\documentclass[sn-mathphys]{sn-jnl}% Math and Physical Sciences Reference Style
\jyear{2022}%
\usepackage{graphicx}% Include figure files
\usepackage{dcolumn}% Align table columns on decimal point
\usepackage{bm}% bold math
\usepackage{hyperref}% add hypertext capabilities
\usepackage[mathlines]{lineno}% Enable numbering of text and display math
\usepackage{xcolor}
\usepackage{booktabs}

\newcommand{\myincludee}[1]{0}
\usepackage{empheq}
\usepackage{siunitx}
\usepackage{amsmath,amsfonts,amssymb}

\newcommand{\dif}{\mathrm{d}}

\usepackage{etoolbox}
\newbool{FastCompilation}
\boolfalse{FastCompilation}
\begin{document}

% \preprint{APS/123-QED}

\title[Thermodynamic optical pressures in tight-binding systems]{Thermodynamic optical pressures in tight-binding nonlinear multimode photonic systems}
% \thanks{A footnote to the article title}%

\author*[1,2]{\fnm{Nikolaos K.} \sur{Efremidis}}\email{nefrem@uoc.gr}

\author[3]{\fnm{Demetrios N.} \sur{Christodoulides}}

% \affil*[1]{\orgdiv{Department}, \orgname{Organization}, \orgaddress{\street{Street}, \city{City}, \postcode{100190}, \state{State}, \country{Country}}}

\affil[1]{\orgdiv{Department of Mathematics and Applied Mathematics}, \orgname{University of Crete},
  \orgaddress{
    % \street{Street},
    \city{Heraklion}, \postcode{70013},
    \state{Crete},
    \country{Greece}}}

\affil[2]{\orgdiv{Institute of Applied and Computational Mathematics}, \orgname{Foundation for Research and Technology-Hellas (FORTH)},
  \orgaddress{
    % \street{Street},
    \city{Heraklion}, \postcode{70013},
    \state{Crete},
    \country{Greece}}}

\affil[3]{\orgdiv{CREOL/College of Optics}, \orgname{University of Central Florida},
  \orgaddress{
    % \street{Street},
    \city{Orlando}, \postcode{32816},
    \state{Florida},
    \country{USA}}}

\date{\today}% It is always \today, today,
             %  but any date may be explicitly specified

\abstract{
  Optical forces are known to arise in a universal fashion in many and diverse physical settings. As such, they are successfully employed over a wide range of applications in areas like biophotonics, optomechanics and integrated optics. While inter-elemental optical forces in few-mode photonic networks have been so far systematically analyzed, little is known, if any, as to how they manifest themselves in highly multimoded optical environments. In this work, by means of statistical mechanics, we formally address this open problem in optically thermalized weakly nonlinear heavily multimode tight-binding networks. The outlined thermodynamic formulation allows one to obtain in an elegant manner analytical results for the exerted thermodynamic pressures in utterly complex arrangements-results that are either computationally intensive or impossible to obtain otherwise. Thus, we derive simple closed-form expressions for the thermodynamic optical pressures displayed among elements, which depend only on the internal energy as well as the coupling coefficients involved. In all cases, our theoretical results are in excellent agreement with numerical computations. Our study may pave the way towards a deeper understanding of these complex processes and could open up avenues in harnessing radiation forces in multimode optomechanical systems.  
}

%\keywords{keyword1, Keyword2, Keyword3, Keyword4}

%\keywords{Suggested keywords}%Use showkeys class option if keyword
                              %display desired
\maketitle

\section*{Introduction}

The advent of Maxwell's electrodynamics was from its very start instrumental in understanding and predicting radiation pressure effects~\cite{lebed-adp1901,nicho-pr1901}. Yet, it was not until the invention of the laser that these processes received widespread attention. The pioneering work of Ashkin and colleagues ushered an era in utilizing optical forces for trapping and manipulating micro-particles and even individual atoms-effects that are nowadays extensively used in biology, biophotonics, soft matter, and atomic physics~\cite{ashki-prl1970,ashki-ol1980,aski-science1980}. In such environments, optical radiation forces arise because of scattering that tends to alter the electromagnetic momentum balance as imposed by the incident wave~\cite{burns-prl1989,burns-science1990,tatar-prl2002}. As such, these forces are typically analyzed by invoking the Maxwell stress tensor formalism once the vectorial electromagnetic field has been determined. In 2005, the prospect for observing a class of radiation forces was suggested~\cite{povin-ol2005,povin-oe2005}. Unlike scattering arrangements, the optical force between two evanescently coupled parallel waveguide elements or optical microcavities can now be either attractive or repulsive depending on the modal excitation conditions~\cite{povin-ol2005,povin-oe2005}. These intriguing evanescent ``optical bonding'' forces were subsequently observed in a series of experimental works involving coupled dielectric waveguides and free-standing lightguides on top of dielectric substrates~\cite{li-nature2008,li-np2009,roels-nn2009}. Over the years, a variety of other experimental setups was also pursued to study optical bonding forces. In~\cite{eiche-np2007,perni-oe2009b}, evanescent coupling from a waveguide to a high-Q whispering-gallery resonator was suggested and utilized. In addition, optical bonding forces have been theoretically and experimentally investigated in vertically coupled ring resonators~\cite{rakic-np2007,rosen-np2009,wiede-nature2009,jiang-oe2009}, in photonic crystal photomechanical cavities~\cite{eiche-nature2009} as well as in coupled microring systems~\cite{pi-acs2022}. As indicated in several studies, the use of such optomechanical forces may be of fundamental importance for integrated photonic applications~\cite{vanth-np2010,metac-apr2014}. Such applications could range from optical routing~\cite{rosen-np2009} and optical information storage~\cite{huang-acs2019}, to precision measurements~\cite{anets-np2009}, photothermal sensing~\cite{prues-acs2018}, and actuators~\cite{li-np2009,roels-nn2009,ren-ACSNano2013}, to mention a few. Along the theoretical front, several works have considered similar situations on the basis of the Maxwell stress tensor and different realizations have been suggested~\cite{mizra-oe2005,rakic-oe2009,perni-oe2009,yang-nl2009,rodri-jn2016,rodri-ol2017,miri-ol2018}.

At this point it is important to note, that so far, research in this area has been exclusively conducted in optomechanical arrangements consisting of only two evanescently coupled optical components such as coupled resonators or waveguides. In this case, the optical pressure is then computed from the very distribution of the optical fields, given that the in-phase supermode leads to attractive forces while the out-of-phase to repulsive interactions between the two elements of the system. Beyond these few-mode optical configurations, there is currently little if any knowledge as to how such forces manifest themselves. Clearly of importance will be to develop pertinent methodologies that are capable of analyzing radiation pressure effects in much more complex settings that could in principle involve hundreds or thousands of modes.

In this work, we investigate for the first time to the best of our knowledge radiation pressure effects in thermalized~\cite{note3} highly-multimoded weakly nonlinear tight binding optical systems. The thermodynamic pressure as exerted collectively by all the modes is obtained using a statistical mechanical formulation~\cite{pathr-2011,callen-1998}, and is analyzed in pertinent photonic discrete systems (comprised of optical waveguides or microresonators) that have attained Rayleigh-Jeans thermal equilibrium conditions. Our formalism is used to study both one- and two-dimensional discrete lattices. These effects are investigated under zero and periodic boundary conditions, corresponding to linear geometric configurations and circular arrays, respectively. In deploying our approach, a grand-canonical description is used, whereby the thermodynamic pressure is defined as the conjugate intensive variable with respect to variations in the separation distance between adjacent elements in the lattice. Importantly, we find that the thermodynamic pressure between the array elements can be elegantly expressed as a function of the optical energy and the underlying coupling coefficients. In two transverse dimensions, the pressure along each direction is, in principle, different. Depending on the internal energy and the ensued optical temperature, the pressure can be positive, negative, or even zero. By using the equilibrated power mode occupancies, our approach can accurately provide the inter-waveguide, or more generally inter-elemental pressures, for different types of boundary conditions. For periodic boundaries, the thermodynamic pressure is equal to all the interelemental pressures. Thus, all the optomechanical forces in a circular array have the same magnitude, which is thermodynamically determined. On the other hand, for zero boundary conditions, the interelemental pressure in general varies along the array. In all cases, the thermodynamic pressure is found to be equal to the average value of the inter-waveguide pressures. Importantly, in our work we provide a complete statistical mechanical formulation of thermalized discrete optical systems, including expressions for the total energy and its differential. A Gibbs-Duhem equation that defines the relation between variations in the intensive parameters of the system is also presented. Our results may be useful in predicting optomechanical forces in utterly complex multicomponent discrete optical settings that could be realized on a variety of integrated photonic platforms. 

\section*{Results}

\subsection*{Optical thermodynamic pressures in one-dimensional nonlinear tight-binding photonic lattices}

\begin{figure*}[h]
  \centering
  \ifbool{FastCompilation}{}{
    \includegraphics[width=\textwidth]{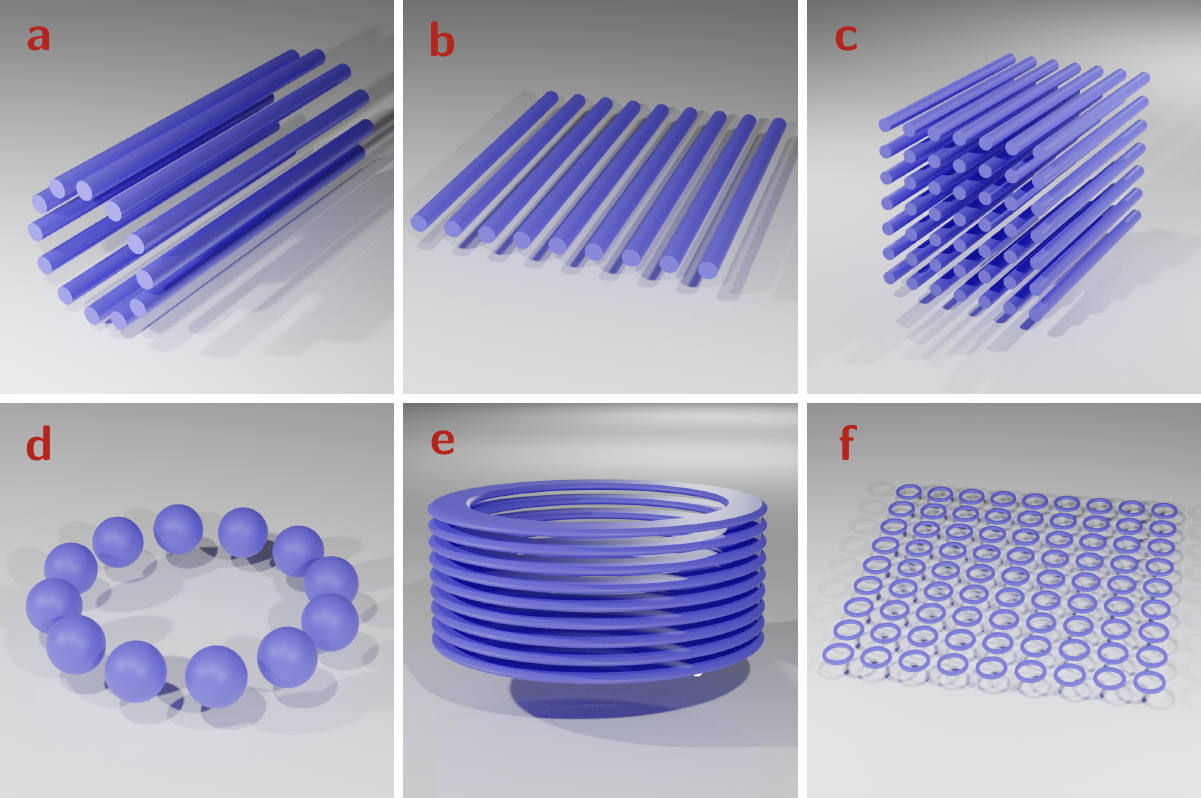}
  }
  \caption{\textbf{Schematic showing typical multimoded optical settings where out theory can be applied to derive a thermodynamic pressure}. A one-dimensional array consisting of evanescently coupled single mode waveguides in \textbf{a} a circular (normal polygon with the wavegguides at the vertices) and \textbf{b} a linear geometric configuration. \textbf{c} A two-dimensional waveguide array forming a rectangular lattice.
    A one-dimensional coupled resonator array in \textbf{d} a circular and \textbf{e} a vertical ring arrangement. \textbf{f} A two-dimensional rectangular lattice of ring resonators.
    In the first column the boundary conditions are periodic, whereas in the second and the third columns the boundary conditions are zero.}
  \label{fig:1}
\end{figure*}
Let us first consider that the system under examination is a, regularly spaced, one-dimensional lattice of waveguides or resonators as shown schematically in the first two columns of Fig.~\ref{fig:1}. For simplicity, each element of the lattice is selected to be single-mode and, thus, the total number of supermodes $M$, is equal to the number of elements or nodes in the lattice. Using a tight-binding approximation~\cite{haus-1984,vahal-2004,fan-josaa2003}, we find that the Hamiltonian of the system satisfies
\[
  H = -\sum_{m=1}^M
  \left[\kappa u_m^*(u_{m+1}+u_{m-1})+\gamma\frac{\lvert u_m\rvert^4}2\right],
\]
where $u_m$ is the nodal amplitude of single-mode element $m$, $\kappa$ is the coupling coefficient between adjacent waveguides, and $\gamma$ determines the strength of the nonlinearity due to the Kerr effect~\cite{leder-pr2008}. The discrete nonlinear Schr\"odinger (DNLS) equation $i\dot u_m+\kappa(u_{m+1}+u_{m-1})+\gamma\lvert u_m\rvert^2u_m=0$ is derived from $\dot u_m=\{H,u_m\}$ by utilizing the Poisson brackets $\{u_m,u_{m'}^*\} = i\delta_{m,m'}$ and $\{u_m,u_{m'}\}=\{u_m^*,u_{m'}^*\}=0$. Such a coupled-mode theory analysis applies in the weakly guided regime~\cite{haus-1984} ($\vert\kappa\vert\ll\omega$ in the temporal domain and $\vert\kappa\vert\ll\beta$ in the spatial domain). In our simulations, both 1D and 2D tight-binding lattice were found to thermalize to the theoretically predicted Rayligh-Jeans distribution. Kinetic nonequilibrium theories can be utilized to theoretically analyze the process of thermalization~\cite{picoz-pr2014,chioc-epl2016}. In a recent study of beam-cleaning in multimode fibers, the condensate fraction and chemical potential are infered from the intensity profiles~\cite{baudi-prl2020}. Using statistical mechanics, the non-trivial task of thermalization to the Rayleigh-Jeans distribution and beam self-cleaning have been observed in multimode fibers~\cite{pourb-np2022,mangi-oe2022}.

Our focus from this point on in this work is going to be concentrated in waveguide arrays, where $\dot u_m = \dif u_m/\dif z$ and $z$ is the propagation distance. However, the same analysis can be directly applied for coupled resonators in which case $\dot u_m=\dif u_m/\dif t$, where $t$ is time. The boundary conditions can either be periodic in the case of a circular configuration, with the nodes being the vertices of a normal polygon [Fig.~\ref{fig:1}a], or zero for a linear geometric arrangement [Fig.~\ref{fig:1}b]. Specifically, the propagation constants of the modes are $\varepsilon^{(l)}=-2\kappa\cos(2\pi l/M)$ for periodic and $\varepsilon^{(l)}=-2\kappa\cos(\pi l/(M+1))$ for zero boundary conditions. Due to the Hermitian nature of the Hamiltonian the respective supermodes are orthogonal. Since the density of states inside the band when $M$ is large is the same in both cases, the boundary conditions do not affect the overall equilibrium thermodynamic behavior of these systems. The propagation constants are linear functions of the coupling coefficient, $\kappa$, which depends on the geometric and index characteristics of the medium. Assuming waveguides with radially symmetric index profile, then $\kappa=\kappa(s)$, where $s$ is the spacing between successive waveguides. For $M\gg1$, the length of the lattice, for both types of boundary conditions, is $L\approx Ms$.

We utilize a recently developed weakly nonlinear optical thermodynamic theory with a supermodal basis~\cite{wu-np2019}. Subsequently, different aspects to this problem have been examined~\cite{makri-ol2020,ramos-prx2020}. Here, in order to be able to define pressure, we follow an approach similar to Ref.~\cite{efrem-pra2021}, that takes into account the additional system parameters and their conjugate variables. For one-dimensional geometries, this additional parameter is the length $L$ of the array. The detailed derivations are presented in the Methods section. We decompose of the optical wave into the supermodes $u_m^{(l)}$ of the lattice, $u_m(z)=\sum_{l=1}^M c^{(l)}(z)u_m^{(l)}$, where $H(u_m^{(l)};\gamma=0) = \varepsilon^{(l)}u_m^{(l)}$, $\varepsilon^{(l)}$ is the eigenvalue or the propagation constant of $u_m^{(l)}$, and $c^{(l)}(z)$ is the respective $z$-dependent amplitude. The optical system conserves the total power $N = \sum\nolimits_ln^{(l)}$,where $n^{(l)}=\lvert c^{(l)}\rvert^2$ is the power occupation number of mode $l$, as well as the Hamiltonian $H$. In the weakly nonlinear regime, we assume that the major contribution to the Hamiltonian originates from the linear part. Thus the total energy per unit length along the longitudinal direction is $U = \sum\nolimits_l\varepsilon^{(l)}n^{(l)}/\omega$, where $\omega$ is the frequency of the electromagnetic wave.

Following the calculations, we find that the power occupation numbers follow a Rayleigh-Jeans distribution $n^{(l)}=1/(\alpha+\beta\varepsilon^{(l)}/\omega)$, where we denote by $\varepsilon$ the set $\varepsilon=\{\varepsilon^{(1)},\ldots,\varepsilon^{(M)}\}$ and $\overline\varepsilon^{(l)}=\varepsilon\setminus\{\varepsilon^{(l)}\}$. In addition the equation of state that relates the internal energy, the power, and the number of modes, with the optical temperature (or temperature) $T$ and the chemical potential $\mu$ is given by $U - \mu N = MT$. Note that the expressions for the distribution and the equation of state are identical to those derived in~\cite{wu-np2019} (where a microcanonical ensemble is utilized) and do not seem to be affected by the presence of the extra parameter $L$. Equivalently, assuming that the number of waveguides of the lattice is fixed, it can be more convenient to define this additional parameter as the spacing between adjacent waveguides $s=L/M$. Importantly, the thermodynamic variable that is conjugate to $L$, is the actual optomechanical electromagnetic pressure that is applied between the waveguides of the lattice. Following the calculation [see Methods section] we can express the pressure in the form
\begin{equation}
  p 
  =
  -
  \sum_l\frac{n^{(l)}}{\omega}
  \left(
    \frac{\partial\varepsilon^{(l)}}{\partial L}
  \right)_{M}.
  \label{eq:p0}
\end{equation}
Since the propagation constants are linearly dependent from the coupling coefficients, the above formula becomes
\begin{equation}
  p
  =
  -\frac UM
  \frac{\dif\log\kappa}{\dif s}.
  \label{eq:p}
\end{equation}
Equation~(\ref{eq:p}) is an surprisingly simple expression that relates the actual thermodynamic pressure with the internal energy and the coupling coefficient by taking into account all the partial pressures exerted by each supermode. Specifically, we see that the pressure depends on two terms: It is proportional to the average internal energy per waveguide $U/M$ and the logarithmic derivative of the coupling coefficient. For a coupling coefficient that decays exponentially $\kappa(s)=\kappa_0e^{-\gamma s}$, the pressure takes the form $p = \gamma U/M$. In general, as long as $\kappa(s)$ is derived analytically, a closed-form expression is obtained for the thermodynamic pressure.

\begin{figure*}[h]
  \centering
  \ifbool{FastCompilation}{}{
    \includegraphics[width=\textwidth]{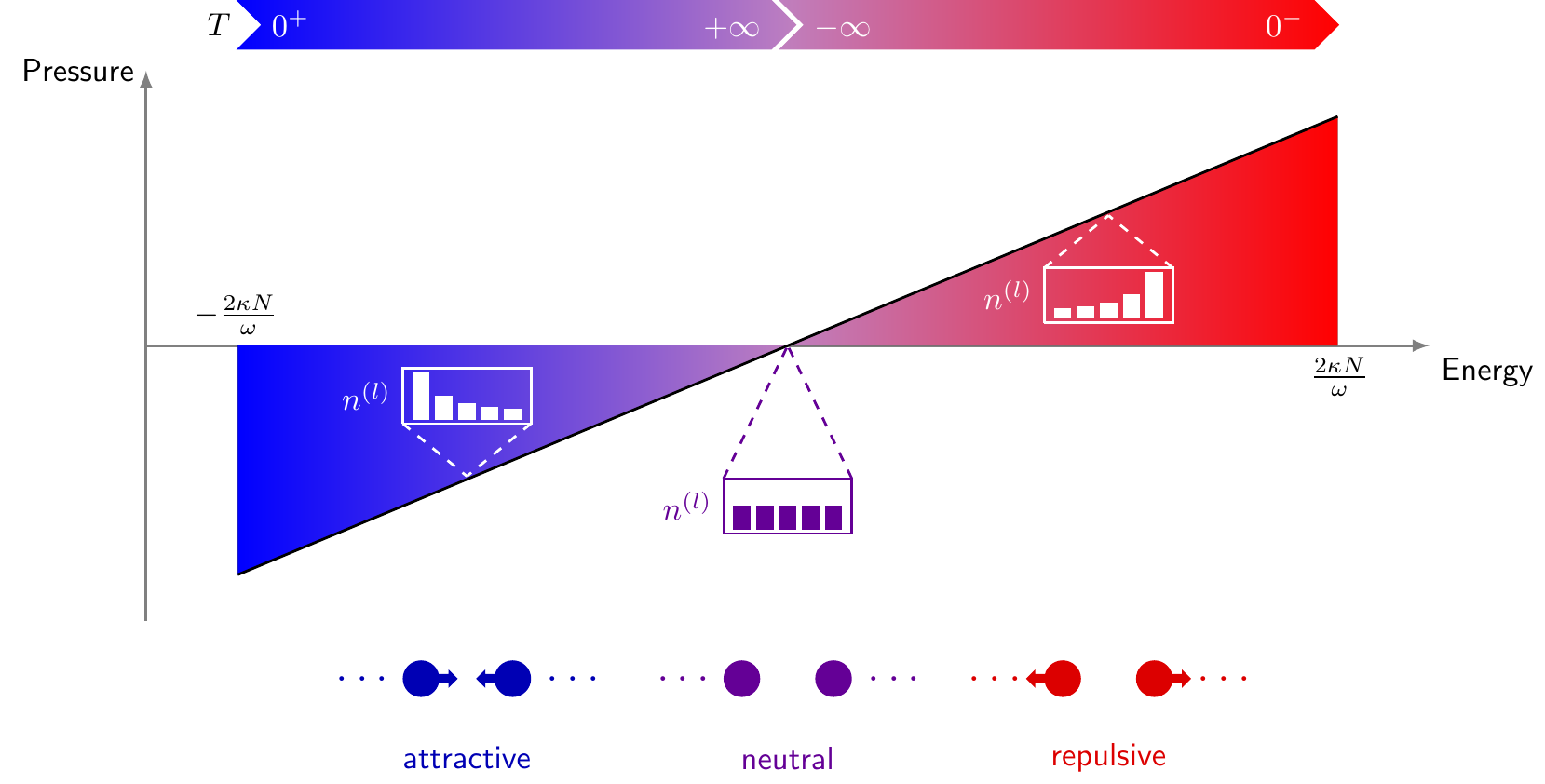}
  }
\caption{\textbf{Relation between the pressure and the internal energy in 1D waveguide arrays.} Schematic showing the linear dependence of the optomechanical pressure from the total or internal energy of the electromagnetic field, under thermal equilibrium conditions [see Eq.~(\ref{eq:p})] in one-dimensional arrays of evanescently coupled waveguides. Lower values of the energy are associated with negative pressure ensuing attractive forces between the waveguides. In addition, the optical temperature is positive and the power modal occupation numbers satisfy a Rayleigh-Jeans distribution that favors the lower order modes. Note that there is a particular value of the energy where the distribution of the power occupation numbers becomes uniform, in which case the temperature is infinite, and the effective pressure between waveguides is zero. For higher values of the energy the pressure is positive, and thus the forces between waveguides become repulsive. The optical temperature is negative and the power modal occupation numbers satisfy a Rayleigh-Jeans distribution that favors the higher order modes. The forces exerted between adjacent waveguides in an array are illustrated at the bottom. 
}
\label{fig:2}
\end{figure*}

From Eq.~(\ref{eq:p}) we can determine the optomechanical pressure, or the force per unit of propagation length that is exerted on the waveguides, due to the gradient optical forces. Since $\dif\kappa/\dif L<0$, Es.~(\ref{eq:p}) predicts that $U$ and $p$ have the same sign (The propagation constants are shifted with respect to the middle of the band. As a result, the total energy can become positive, zero, or negative). Positive (negative) pressure is associated with a repulsive (attractive) force between adjacent waveguides. In the low and high energy condensation limits the system approaches the in-phase and out-of-phase modes. The respective energies, temperatures, and pressures can be analytically computed and are given by $U=-2N\kappa/\omega$, $T=0^+$, $p=(2N/(\omega M))\dif\kappa/\dif s$ and $U=2N\kappa/\omega$, $T=0^-$, $p=-(2N/(\omega M))\dif\kappa/\dif s$. In the middle of the band $U=0$, $T\rightarrow\pm\infty$ and the average pressure is zero, $p=0$. As we can see in Fig.~\ref{fig:2}, according to the Rayleigh-Jeans distribution, when $U<0$ the lower-order modes have higher occupation numbers. The distribution is inverted when $U>0$ resulting to larger occupation numbers for the higher-order modes. Taking into account the phase profile of the supermodes, we can say that the array is biased towards an in-phase (out-of-phase) structure when $U<0$ ($U>0$), leading to attractive (repulsive) forces, respectively. In Fig.~\ref{fig:2}, we summarize our results concerning the pressure that is exerted between two adjacent waveguides in a one-dimensional waveguide array. In particular, we depict the variations in the pressure and the temperature as a function of the total energy, as well as the associated power modal occupation numbers. The forces exerted between adjacent waveguides are schematically shown. The total force on a waveguide is determined by vectorially adding all the interwaveguide pressures.

In addition to pressure, we have derived expressions that determine the thermodynamic state of an optically thermalized (or thermalized) system. In particular, our calculations show that
\begin{equation}
  \dif U = T\dif S - pM\dif s+\mu\dif N-R\dif M,
  \label{eq:dU}
\end{equation}
a formula that shows how the internal energy is modified due to variations in the spacing between waveguides, the power, or the number of modes of the system. In Eq.~(\ref{eq:dU}), $R = T (  \partial q/\partial M )_{\mu,T,s}$ is the conjugate variable to the number of modes and $q$ is the $q$-potential that is directly related to the grand-canonical partition function [see Methods]. In the case of one transverse dimension, waveguide arrays constitute an extensive system~\cite{efrem-pra2021,wu-np2019} meaning that the entropy $S$, is a homogeneous function of degree one, with respect to $U$, $M$, $N$, or $S(\lambda U,\lambda M,\lambda N) = \lambda S(U,M,N)$. As a result, we are able to integrate the extensive variables while keeping the intensive quantities constant in Eq.~(\ref{eq:dU}) leading to the following relation for the total or internal energy
\begin{equation}
  U = TS+\mu N-RM.
  \label{eq:U}
\end{equation}
We see that we can decompose $U$ into three terms involving the entropy, the power, and the number of modes, with the respective conjugate variables being $T$, $\mu$, and $R$. Note that in classical thermodynamics the energy decomposition contains a term that is proportional to the pressure. However, here this is not the case. In particular, the pressure-volume term is replaced with the $RM$ product. To highlight the similarity of $R$ with the pressure of classical thermodynamics, we define $R$ as the ``internal pressure''. The internal pressure physically describes how much the internal energy is modified when an additional mode is introduced to the system. However, we are not able to see any observable physical significance of this term in discrete photonic systems. Another consequence of the extensive character of the entropy is the following simplified expression for the internal pressure $R=qT/M$. Finally, from Eqs.~(\ref{eq:dU}), (\ref{eq:U}), we derive the Gibbs-Duhem equation~\cite{callen-1998}
\begin{equation}
  S\dif T +pM\dif s+N\dif\mu -M\dif R = 0,
\end{equation}
which shows that the four intensive parameters $T$, $s$, $\mu$, and $R$ are interdependent.

\subsection*{Radiation pressures in thermalized two-dimensional discrete array systems}
Our results can be generalized in the case of two-dimensional evanescently coupled discrete optical configurations. Here, for simplicity, we assume a two-dimensional rectangular arrangement of single-mode waveguides [Fig.~\ref{fig:1}c] and account for coupling between first neighbors. Thus, the Hamiltonian of the system takes the form
\[
H = -\sum_m
\left[
  u_m^*(\kappa_1\Delta_1u_m+\kappa_2\Delta_2u_m)+\gamma\frac{\lvert u_m\rvert^4}2
\right],
\]
where $m=(m_1,m_2)$, $m_j=1,\ldots,M_j$, with $j=1,2$, $\Delta_1u_m=u_{m_1+1,m_2}+u_{m_1-1,m_2}$, $\Delta_2u_m=u_{m_1,m_2+1}+u_{m_1,m_2-1}$, and $M=M_1M_2$. The spectrum of the linear modes is given by $\varepsilon^{(l)}=\varepsilon_0^{(l_1)}(M_1,s_1)+\varepsilon_0^{(l_2)}(M_2,s_2)$, where $\varepsilon_0^{(l_j)}(M_j,s_j)=-2\kappa_j\cos(\pi l_j/(M_j+1))$, $l=(l_1,l_2)$, $\kappa_j=\kappa(s_j)$ is the coupling coefficient and $s_j$ the distance between adjacent elements of the lattice along the $j$th direction. Assuming that the major contribution to the Hamiltonian arises from the linear terms, then $U=U_1+U_2$, where $U_j=\sum_l\varepsilon^{(l_j)}_0n^{(l)}/\omega$ is the part of the energy per unit of propagation length associated with coupling along the $j$th transverse direction. We can express the two energy components as $U_j=M\kappa_j\mathcal U_j$. Importantly, we can define the electromagnetic pressure along the two transverse directions as $p_j=-\sum_{j=1}^2\sum_{l=1}^{M}(n^{(l)}/\omega)(\partial\varepsilon_0^{(l_j)}/\partial A)_{M_1,M_2,s_{3-j}},$ where $A=M_1M_2d_1d_2$ is the transverse area occupied by the lattice, and $j=1,2$ are the $x$, and $y$ directions, respectively. Then, for example, the pressure $p_1=p_x$ describes the applied force per unit area in the $y-z$ plane. The above expression can be simplified due to the linear dependence of the propagation constants from the coupling coefficients as
\begin{equation}
  p_j
  =
  -\frac{U_j}{M}\frac{1}{s_{3-j}}
  \frac{\dif\log\kappa_j}{\dif s_j}. 
  \label{eq:p2d}
\end{equation}
This formula is a direct generalization of Eq.~(\ref{eq:p}). We see that the pressure is proportional to two terms: The first is the total energy due to coupling along the $j$th direction per waveguide, while the second is the logarithmic derivative of the coupling coefficient with respect to the area of the primitive cell $s_1s_2$ due to variations along the $j$th direction. From Eq.~(\ref{eq:p2d}) we see that the pressures along each direction are, in general, different depending for example on the spacing $s_j$. In addition, we find that the differential of the internal energy is given by
\begin{equation}
  \dif U = T\dif S +\mu\dif N-p_1Ms_2\dif s_1-p_2Ms_1\dif s_2
  - 
  R_1M_2\dif M_1-R_2M_1\dif M_2,
  \label{eq:difU2D}
\end{equation}
where the internal pressure is 
$R_j = (T/M_{3-j})(\partial q/\partial M_j)_{M_{3-j},\mu,T,s}$.

Note that Eq.~(\ref{eq:difU2D}), in general, can not be integrated. For example, when the number of waveguides along one direction is small, even if the total number of waveguides $M$ is large, the array exhibits non-extensive corrections. It is only in the limit where both $M_1$, $M_2$ are large enough that the system asymptotically behaves as extensive, meaning that $S(\lambda U,\lambda N,\lambda M)=\lambda S(U,N,M)$ (see a detailed discussion in~\cite{efrem-pra2021}). There are several consequences of extensivity that simplify the resulting thermodynamic description. In particular, when the system reaches thermal equilibrium, due to equidistribution of the energy, and in agreement with our simulations, $\mathcal U_1=\mathcal U_2=\mathcal U$ and, thus, the total energy can be written as $U=(\kappa_1+\kappa_2)M\mathcal U$. In addition the internal pressures along each direction are equalized $R=R_1=R_2=qT/M$, and the differential of the internal energy is written as 
\begin{equation}
  \dif U = T\dif S +\mu\dif N-p_1Ms_2\dif s_1-p_2Ms_1\dif s_2
  - 
  R\dif M.
  \label{eq:dU2d}
\end{equation}
Equation~(\ref{eq:dU2d}) can then be directly integrated to $U = TS+\mu N -RM$. Combining the previous expressions, we derive a Gibbs-Duhem equation 
\begin{equation}
  S\dif T +N\dif\mu +p_1Ms_2\dif s_1+p_2Ms_1\dif s_2-M\dif R = 0 
\end{equation}
that relates the intensive variables of the system. We can express the pressure along the $j$th direction in terms of $\mathcal U$ as $p_j =-(\mathcal U/s_{3-j})(\dif\kappa_j/\dif s_j).$ However, it is physically relevant to also define the force per unit of propagation length that is exerted to a single waveguide along the $j$th direction $\hat p_j$ (rather than the force per unit of transverse area). This is simply the product $p_js_{3_j}$ or
\begin{equation}
  \hat p_j = 
  -\mathcal U
  \frac{\dif\kappa_j}{\dif s_j}.
\end{equation}
Thus, for an extensive system, the pressure is the same along the two transverse directions only if the lattice is square ($s_1=s_2$). In addition, since the sign of the pressure depends on the sign of $\mathcal U$, the pressures along both direction should always have the same sign. 

Below, we analyze the pressure distributions under zero and periodic boundary conditions. In both cases the average value of the pressure is equal to the thermodynamic pressure. Furthermore, in the case of periodic boundary conditions, the magnitude of the interwaveguide pressures are all equal to the thermodynamic pressure. The differences between these two types of boundary conditions are associated with the symmetries of the system and are rendered in the respective modal distributions.

\subsection*{Inter-waveguide thermodynamic optical pressures in tight-binding lattices\label{sec:interwaveguide}}

In the remaining part of this paper, we will investigate the effect of boundary conditions on the actual pressure that is exerted between all adjacent waveguides of a lattice. We will focus in the case of one-dimensional lattices, although the same principles apply for two-dimensional configurations. Importantly, we are going to determine how the thermodynamic pressure given by Eq.~(\ref{eq:p}) compares to these interwaveguide pressures $p_j$, between the waveguide elements $j$ and $j+1$, for both zero and periodic (Born-von Karman type) boundary conditions. 

We directly compute the interwaveguide pressure $p_j$, by decomposing the lattice into a set of virtual directional couplers. For each one of these couplers, we apply the results of Ref.~\cite{rakic-oe2009}. Adapting these expressions to our formulation, we determine that the pressure for the two modes of the $j$th coupler is given by 
\begin{equation}
  p_{j,\pm}=\pm\frac{N_{j,\pm}}{\omega}\frac{\dif\kappa}{\dif s},
  \label{eq:rakich}
\end{equation}
where the plus (minus) sign accounts for the in-phase (out-of-phase) mode, and $N_{j,\pm}$ is the respective power. Any arbitrary amplitude profile of a virtual coupler can then be written as a superposition of these two modes. By adding the partial pressures, we compute the total pressure $p_j=p_{j,+}+p_{j,-}$ between waveguides $j$ and $j+1$. Note that since we are utilizing a statistical mechanical formulation, $p_j$, $j=1,\ldots,M-1$, are actually the average values of an ensemble under thermal equilibrium.

Note that configurations, such as those shown in Fig.~\ref{fig:1}, can be utilized to observe optomechanical forces. In order for the waveguides or resonators to move almost freely, techniques used in two-component systems~\cite{li-nature2008,li-np2009,roels-nn2009,eiche-np2007,perni-oe2009b, rakic-np2007,rosen-np2009,wiede-nature2009,jiang-oe2009,eiche-nature2009,pi-acs2022} can be adapted to minimize mechanical stiffness without sacrificing mechanical stability. For example, the waveguide arrays depicted in the first row of Fig.~\ref{fig:1} might be suspended and clamped from both sides.

\subsection*{Inter-waveguide thermodynamic optical pressures under zero-boundary conditions}

When the boundary conditions of the system are zero [see the first two columns of  Fig.~\ref{fig:1}], then the amplitude profile of the modes can be highly anisotropic. Specifically, in the case of one-dimensional lattices which are terminated on both sides, i.e. $u_0=u_{M+1}=0$, the supermodes are given by $u_m^{(l)}=(2/(M+1))^{1/2}\sin(\pi l m/(M+1))$. The effect of these boundaries is more prominent close to the condensation limits, where mainly the $u_m^{(1)}$ (or the $u_m^{(M)}$) mode is excited. We have numerically computed the pressure distribution along the array $p_j$, as well as the thermodynamic pressure $p$ from Eq.~(\ref{eq:p}). We have limited ourselves to negative and zero energies $U\le0$ or equivalently to $T\ge0$. The case $U>0$ can be trivially obtained from $U<0$ through the transformations $U\rightarrow-U$, $T\rightarrow-T$, $p\rightarrow-p$, $p_j\rightarrow-p_j$, $n^{(l)}\rightarrow n^{(M-l+1)}$.

\begin{figure*}[h]
  \centering
  \ifbool{FastCompilation}{}{
    \includegraphics[width=\textwidth]{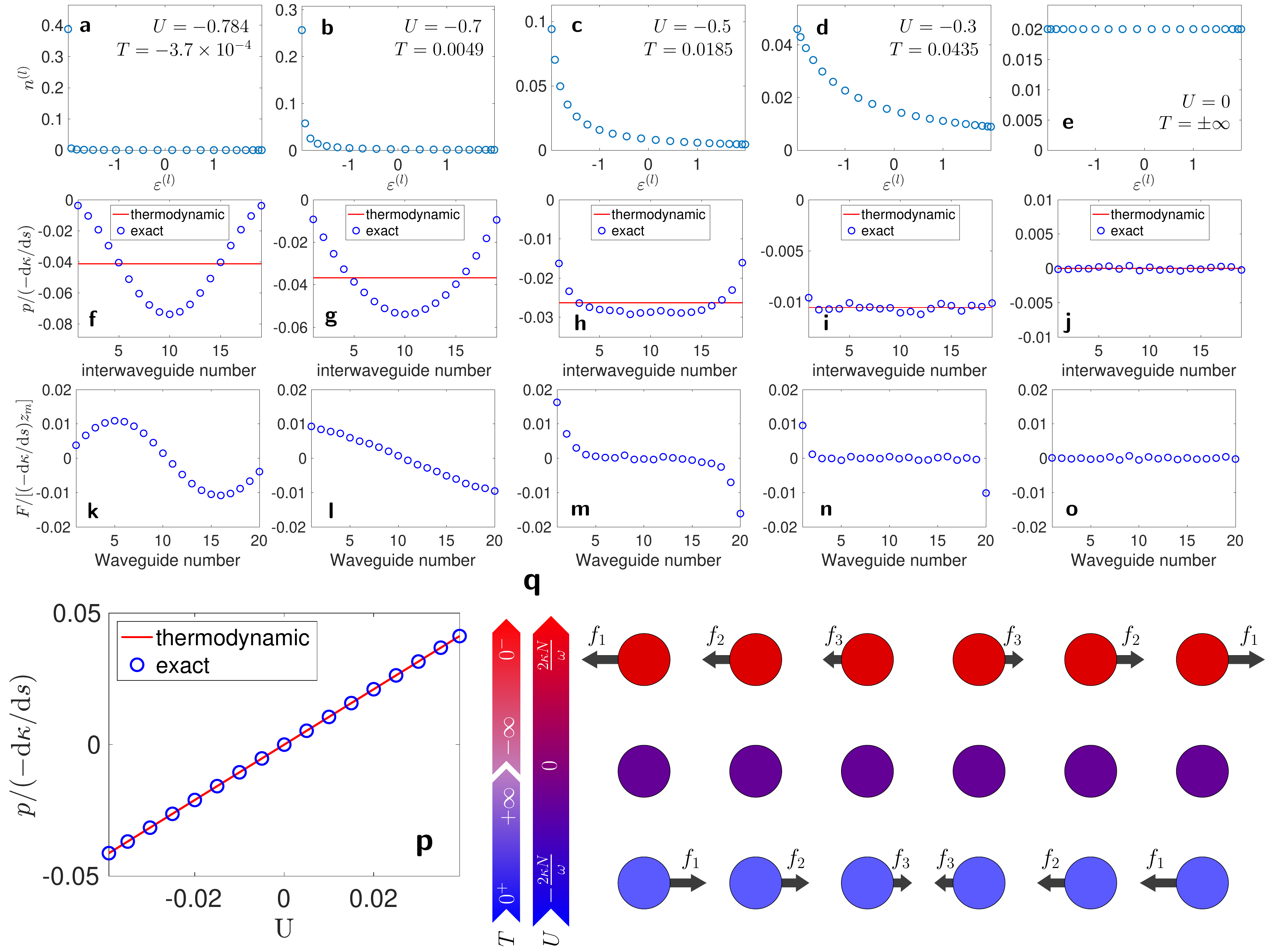}
  }
  \caption{\textbf{Pressure in a linear arrangement of waveguides having zero boundary conditions.}
    Distributions of pressure and applied forces in a one-dimensional photonic lattice with $M=20$ waveguides, zero boundary conditions, and power $N=0.4$. 
    In the first row (\textbf{a}-\textbf{e}) we see the power occupation numbers which obey Rayleigh-Jeans distributions. In the second row (\textbf{f-j}), the interwaveguide pressures $p_j$, obtained by taking an ensemble average (blue circles), and the thermodynamic pressure (red line) are presented and compared. In the third row (\textbf{k-o}) we depict the total force per unit of propagation length applied to each particular waveguide of the lattice.
     In the five columns the internal energy is increased from the condensation limit (first column) to $U=0$ where the Rayleigh-Jeans distribution becomes uniform (as $T\rightarrow\pm\infty$).
     In the first three rows, we focus in the case of negative energies. Results with $U>0$ can be directly derived from those with negative $U$ via a simple transformation, as described in the main text.
     \textbf{p} The thermodynamic pressure is in excellent agreement with the average value of the intewaveguide pressures $\overline p=(1/(M-1))\sum_{j=1}^{M-1}p_j$.
     \textbf{q} Schematic of the forces exerted within the lattice. When $U<0$ or $T>0$ (top) the forces are attractive, whereas when $U>0$ or $T<0$ (bottom) the forces are repulsive. }
  \label{fig:3}
\end{figure*}

A series of numerical results, as obtained by directly solving the coupled-mode theory equations, are shown [Fig.~\ref{fig:3}a-o]. We use normalized units and set $\omega=\kappa=1$. Since we want our results to be independent from the specific waveguide system that we use, we chose to measure the thermodynamic pressure in the form $p/(-\dif\kappa/\dif s)$ [and respectively the interwaveguide pressures as $p_j/(-\dif\kappa/\dif s)$]. We use an array with $M=20$ waveguides, and select the input power to be $N=0.4$. Thus, the internal energy can take values in the range $-0.8\le U\le0.8$. In the three rows of Fig.~\ref{fig:3} we can see the distribution of the power occupation numbers, the interwaveguide pressures, and the total force per unit of propagation length applied in each waveguide. The latter is computed by subtracting the respective left and the right pressures $F_j /[(-\dif\kappa/\dif s)z_m] = (p_{j-1}-p_{j})/(-\dif\kappa/\dif s)$, where $z_m$ is the length of the array. 

In the first column of Fig.~\ref{fig:3}, $U=-0.784$, and thus we are very close to the condensation limit. Almost all of the power relaxes to the lowest order mode of the array. The  pressure between waveguides is stronger at the center of the array as compared to the edges, and the resulting actual forces per unit of propagation length tend to compress the array. Specifically, the attractive forces become maximum close to $n=5$ and $n=15$, and are minimized at the center and the edges of the lattice. In the second column, we increase the energy to $U=-0.7$, a value which is still close to the condensation limit. The main difference here is that the force as a function of the waveguide number tends to behave in a quasi-linear fashion. Thus the forces are maximum at the edges of the array and minimum close to the center. In the third and forth column $U=-0.5$ and $U=-0.3$, respectively, and the Rayleigh-Jeans distribution results to power occupation numbers that are non-negligible for all the modes of the system. When $U=-0.5$ the thermodynamic pressure is almost constant along the array except close to the edges, where the pressure is reduced (in terms of absolute values). This leads to almost zero forces in the bulk of the array, and significant attractive forces at the edges (mainly on the first two and last two waveguides). When $U=-0.3$ the pressure is almost constant everywhere in the array, and thus attractive forces are effectively applied only on the first and the last waveguide of the lattice. Finally, when the internal energy is increased to $U=0$, the modal distribution becomes uniform, the pressure is zero on average, and thus all the net forces on the array are zero. The red lines in the second row of Fig.~\ref{fig:3} are the values of the thermodynamic pressure obtained from Eq.~(\ref{eq:p}). As we can see in Fig.~\ref{fig:3}q the thermodynamic pressure is in excellent agreement with the average value of the interwaveguide pressures $\overline p=(1/(M-1))\sum_{j=1}^{M-1} p_j$. In Fig.~\ref{fig:3}p we can see a schematic of the applied forces along the array. When $U<0$ the forces tend to compress the array, whereas, when $U>0$ the forces are repulsive. Except very close to the condensation limit, the absolute value of the forces are significantly increased as we approach the edges.

\begin{figure*}[h]
  \centering
  \ifbool{FastCompilation}{}{
    \includegraphics[width=\textwidth]{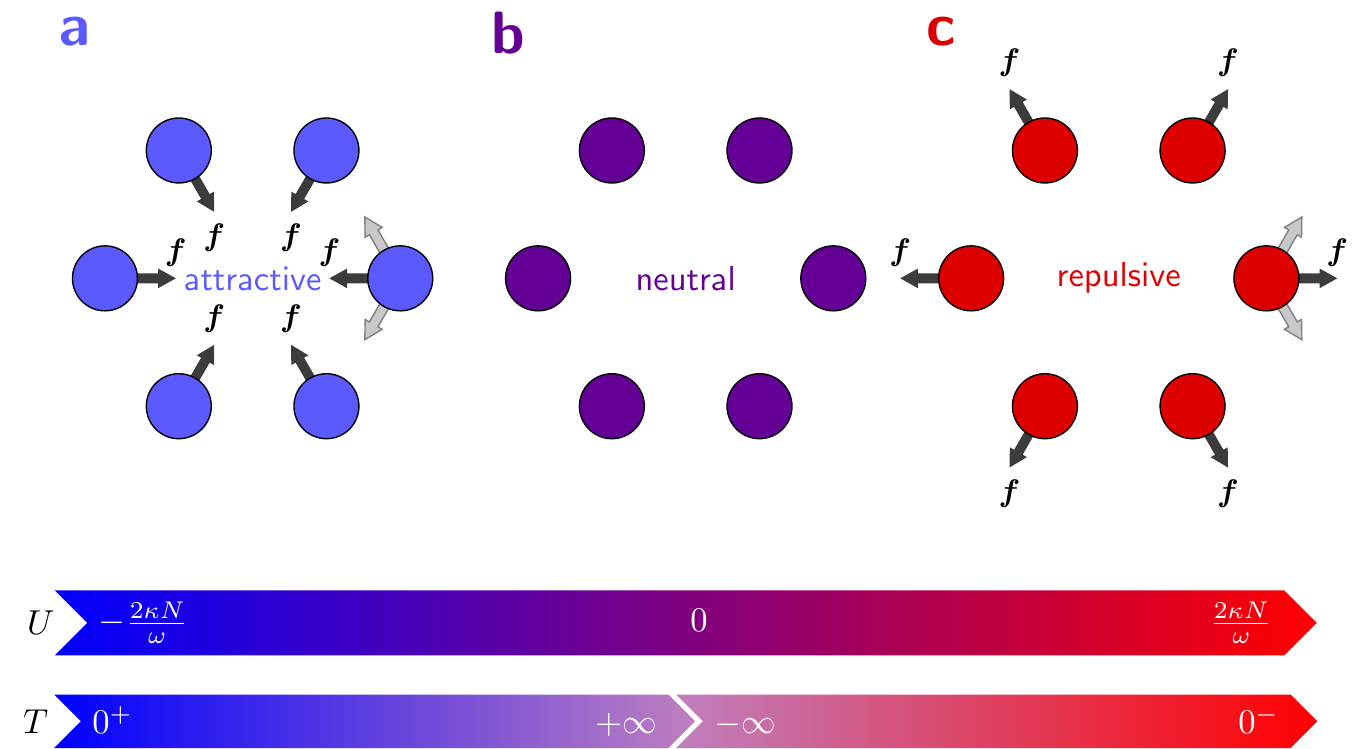}
  }
  \caption{
    \textbf{Pressure in a circular arrangement of waveguides having periodic boundary conditions.}
    \textbf{a} Attractive, \textbf{b} zero, and \textbf{c} repulsive forces in circular discrete optical systems.
    The light gray arrows (shown indicative only on the waveguide located to the right) represent forces exerted due to the interwaveguide pressures. The vectorial sum of these forces in each waveguide are depicted with black arrows. When $U<0$ and thus $T>0$ the forces are attractive, whereas when $U>0$ and thus $T<0$ the forces are repulsive. In between, we have $U=0$ and $T=\pm\infty$ leading to zero net interwaveguide forces.}
  \label{fig:4}
\end{figure*}

\subsection*{Radiation pressures in thermalized suspended circular arrays: periodic boundary conditions}

When the boundary conditions are periodic, then due to symmetry, the statistical properties of all the elements of the lattice are identical. As a result, the average value of the interwaveguide pressure $p_j$ is independent of $j$ and equal to the thermodynamic pressure $p_j=p,\quad j=1,\ldots,M-1$, as confirmed by direct numerical simulations. Thus, without the need for additional calculations, from the thermodynamic pressure given by Eq.~(\ref{eq:p}) we can determine all the interwaveguide forces in the array. In this section, we are going to present a detailed example pertaining to a suspended circular array using physically relevant parameters. 

We assume an evanescently coupling waveguide array in a circular configuration, with the waveguide elements being the vertices of a regular $M$-sided polygon, as shown in Fig.~\ref{fig:4}.  The array is suspended in an index matching fluid and double clamped at $z=0$ and at $z=z_m$. The applied forces are going to displace each waveguide, with the maximum displacement observed at the center of the suspended section $z=z_m/2$, (see for example Ref.~\cite{povin-ol2005}). The calculations for the coupling coefficient between such weakly guided circular waveguides are presented in the Supplementary Note. For a normal polygon with $M$ sides, the vectorial sum of the two forces per unit of propagation length exerted in a single waveguide is equal to
\begin{equation}
  \bm f = \frac{\bm F}{z_m}=2p\sin\frac\pi M\hat{\bm e}_r.
\end{equation}
When $M\gg1$ the above expression takes the form $\bm f \approx (2\pi p/M)\hat{\bm e}_r$. 

As an example, we consider a circular array with core index $n_\mathrm{co}=1.5$ suspended in an index matching fluid with $n_\mathrm{cl}=1.49$. The radius of each core is $\rho=3.2\ \unit{\micro\meter}$, and the distance between successive elements of the lattice is $s=3\rho=9.6$ \unit{\micro\meter}. On average, each single-mode waveguide is loaded with $N/M=1\ \unit{\kilo\watt}$ of power, while the system is operated at a wavelength of $\lambda=1.55$ \unit{\micro\meter}. A number of $M=20$ waveguides is more than sufficient for a thermodynamic description to be applicable~\cite{efrem-pra2021}. Depending on the temperature, or, equivalently on the internal energy, the pressure can take values in the range $-0.587\ \unit{\nano\newton\per\micro\meter}\le p\le0.587 \ \unit{\nano\newton\per\micro\meter}$. Close to these two limiting values of the pressure we reach the condensation limits. Taking $p=-0.3$ \unit{\nano\newton\per\micro\meter}, a value far from condensation, we find that the total force applied in each waveguide is $\bm f=-0.0938\hat{\bm e}_r$ \unit{\nano\newton\per\micro\meter}.

\section*{Conclusion}

In conclusion, following a coupled-mode theory approach, we have derived a complete thermodynamic description for discrete optical systems, such as waveguide arrays and coupled microresonators in one- and two-dimensional arrangements, in the weakly nonlinear regime. Focusing in the case of arrays of evanescently coupled waveguides, we have computed the interwaveguide pressures exerted between adjacent elements of the lattice, for both zero and periodic boundary conditions. Importantly, we have found compact closed-form expressions for the average thermodynamic pressure. The formalism developed here can also be utilized in effectively any physical setting where optical thermodynamics is applicable. These include polyatomic, topological, and non-Hermitian lattices, as well as tight-binding configurations where each element is multimodal. We expect that our results might be useful in predicting optomechanical forces in integrated photonic systems.

\section*{Methods}
%%%%%%%%%%%%%%%%%%%%%
\subsection*{Calculations for one-dimensional photonic lattices}

In what follows, we use a grand-canonical approach similar to Ref.~\cite{efrem-pra2021}, that takes into account the additional system parameters and their conjugate variables. For one-dimensional geometries, this parameter is the spacing between adjacent waveguides $s$, or equivalently, the total length of the array $L$. We expand the optical wave in modal space as
\[
  u_m(z)=\sum_{l=1}^M c^{(l)}(z)u_m^{(l)},
\]
where $H(u_m^{(l)};\gamma=0) = \varepsilon^{(l)}u_m^{(l)}$, $u_m^{(l)}$ is the linear eigenmode with index $l$ and propagation constant or eigenvalue $\varepsilon^{(l)}$, and $c^{(l)}(z)$ is the respective $z$-dependent amplitude. The DNLS equation conserves the total power
\[
  N = \sum\nolimits_ln^{(l)},
\]
where $n^{(l)}=\lvert c^{(l)}\rvert^2$ is the power occupation number of mode $l$, as well as the Hamiltonian $H$. In the weakly nonlinear regime, we assume that the major contribution to the Hamiltonian originates from the linear part~\cite{wu-np2019}. Thus the total energy per unit length along the longitudinal direction is 
\begin{equation}
  U = \sum\nolimits_l\frac{\varepsilon^{(l)}n^{(l)}}{\omega},
\end{equation}
where $\omega$ is the frequency of the electromagnetic wave.

The probability of an optical wave with power occupation numbers $\{n^{(1)},n^{(2)},\ldots\}$ is calculated by minimizing the entropy subject to the two conservation laws~\cite{pathr-2011}
\[
  \rho =
  \exp\left(
    -q-\alpha\sum_ln^{(l)}-\frac\beta\omega\sum_l\varepsilon^{(l)}n^{(l)}
  \right),
\]
where $\alpha$, $\beta$ are the Lagrange multipliers,
\[
  q=\log\mathcal Q
\]
is the $q$-potential, and
\[
  \mathcal Q=\int\prod_{l=1}^M\rho\dif n^{(l)}
  =\prod_{l=1}^M\frac1{\alpha+\beta\varepsilon^{(l)}/\omega}
\]
is the grand canonical partition function. Since the propagation constants are functions of $M$ and $\kappa(s)$, we can express the $q$-potential as $q=q(\alpha,\beta,M;\varepsilon(M,s))$. From the partial derivatives of $q$, we find that the power occupation numbers follow a Rayleigh-Jeans distribution~\cite{wu-np2019}
\begin{equation}
  \langle n^{(l)}\rangle=
  -
  \frac\omega\beta\left(
    \frac{\partial q}{\partial\varepsilon^{(l)}}
  \right)_{\alpha,\beta,M,\overline\varepsilon^{(l)}}
  =
  \frac1{\alpha+\beta\varepsilon^{(l)}/\omega},
  \label{eq:app:nl:nl}
\end{equation}
where we denote by $\varepsilon$ the set $\varepsilon=\{\varepsilon^{(1)},\ldots,\varepsilon^{(M)}\}$ and $\overline\varepsilon^{(l)}=\varepsilon\setminus\{\varepsilon^{(l)}\}$. Furthermore
\begin{equation}
  \langle N\rangle =
  -
  \left(
    \frac{\partial q}{\partial\alpha}
  \right)_{\beta,M,s}
  = \sum_{l=1}^M\langle n^{(l)}\rangle,
  \label{eq:app:N:N}
\end{equation}
\begin{equation}
  \langle U\rangle =
  -
  \left(
    \frac{\partial q}{\partial\beta}
  \right)_{\alpha,M,s}
  =
  \frac1\omega
  \sum_{l=1}^M\langle n^{(l)}\rangle \varepsilon^{(l)},
  \label{eq:app:U:U}
\end{equation}
and
\[
  \mathcal Q = \prod_{l=1}^M\langle n^{(l)}\rangle.
\]
From this point on, we will omit the bracket notation ($\langle\cdots\rangle$) that denotes ensemble averaging. Substituting Eq.~(\ref{eq:app:nl:nl}) to Eqs.~(\ref{eq:app:N:N})-(\ref{eq:app:U:U}), we can derive the following equation of state
\begin{equation}
  \alpha N+\beta U = M.
  \label{eq:app:state:state}
\end{equation}
Taking the differential of $q$
\[
  \dif q =
  \left(
    \frac{\partial q}{\partial\alpha}
  \right)_{\beta,s,M}\dif\alpha
  +
  \left(
    \frac{\partial q}{\partial\beta}
  \right)_{\alpha,s,M}\dif\beta
  +
    \left(
    \frac{\partial q}{\partial s}
  \right)_{\alpha,\beta,M}\dif s
  +
  \left(
    \frac{\partial q}{\partial M}
  \right)_{\alpha,\beta,s}\dif M
\]
and following the calculations one can show that
\[
  \dif(q+\alpha N+\beta U) =
  \beta\left[
    \frac\alpha\beta\dif N +\dif U+pM\dif s+R\dif M
  \right]. 
\]
By interpreting the above equation as the equivalent to the first law of thermodynamics, we find that the Lagrange multiplies can be written in terms of an optical temperature $T$ and a chemical potential $\mu$ as $\beta=1/T$ and $\alpha=-\mu/T$. Thus Eq.~(\ref{eq:app:state:state}) becomes~\cite{wu-np2019}
\begin{equation}
  U - \mu N = MT,
\end{equation}
an expression that relates the two conservation laws and the number of modes with the optical temperature and the chemical potential. In addition, we find that the entropy is associated to the $q$ potential via
\begin{equation}
  S = q+\alpha N+\beta U= q+M,
  \label{eq:app:S}
\end{equation}
while, the differential of $U$ is given by 
\begin{equation}
  \dif U = T\dif S - pM\dif s+\mu\dif N-R\dif M.
  \label{eq:app:dU}
\end{equation}
In Eq.~(\ref{eq:app:dU}) we have defined
\begin{equation}
  R =
  T
  \left(
    \frac{\partial q}{\partial M}
  \right)_{\mu,T,s}
  \label{eq:app:R1}
\end{equation}
to be the conjugate variable to the number of modes.

We define the pressure as the conjugate variable to the length $L=Ms$ of the array 
\begin{equation}
  p = T
  \left(
    \frac{\partial q}{\partial L}
  \right)_{\mu,T,M}
  =
  -
  \sum_l\frac{n^{(l)}}{\omega M}
  \left(
    \frac{\partial\varepsilon^{(l)}}{\partial s}
  \right)_{M}. 
\end{equation}
Note that this is a different definition, in comparison to all the previous works (see for example~\cite{wu-np2019,efrem-pra2021}), where the pressure was considered to be the variable that is conjugate to the number of modes $M$ and thus determined by Eq.~(\ref{eq:app:R1}). Since the propagation constants are linearly dependent from the coupling coefficients, the above formula takes the form
\begin{equation}
  p
  =
  -\frac U{\kappa M}
  \frac{\dif\kappa}{\dif s}.
  \label{eq:app:p}
\end{equation}
Assuming that the coupling coefficient decays exponentially with the distance between adjacent waveguides
\begin{equation}
  \kappa(s)=\kappa_0e^{-\gamma s},
  \label{eq:app:kappa}
\end{equation}
we obtain
\begin{equation}
  p = \gamma\kappa\mathcal U,
  \label{eq:app:p_v2}
\end{equation}
where $\mathcal U=U/(M\kappa)$ is the total energy per waveguide divided with the coupling coefficient. Interestingly, if we follow a similar grand canonical approach and ignore the dependence $\kappa(s)$, i.e., $q=q(\alpha,\beta,M;\varepsilon(M,\kappa))$, then $\mathcal U$ is found to represent a normalized expression for the stress.

In regular one-dimensional tight-binding photonic lattices, the dynamics of the optical wave in the presence of Kerr nonlinearity satisfies the discrete nonlinear Schr\"odinger equation
\[
  i\dot u_m+\kappa(u_{m+1}+u_{m-1})+\gamma\vert u_m\vert^2u_m=0,
\]
where $\gamma$ is the Kerr nonlinear coefficient, and $\kappa$ is the coupling coefficient. In the linear limit with periodic boundary conditions the modes (or supermodes) of the system are
\[
  u_m^{(l)}=
  \frac1{\sqrt{M}}
  \exp\left[
    i\frac{2\pi km}{M}
  \right],
\]
with eigenvalues
\[
  \varepsilon^{(l)}=
  -2\kappa\cos\frac{2\pi l}{M}.
\]
On the other hand, for zero boundary conditions, the modes
\[
  u_m^{(l)}=
  \left(
    \frac{2}{M+1}
  \right)^{1/2}
  \sin
  \frac{\pi lm}{M+1}
\]
are associated with the eigenvalues
\[
  \varepsilon^{(l)}
  =
  -2\kappa\cos
  \frac{\pi l}{M+1}.
\]
%}
%%%%%%%%%%%%%%%%%%%%%

\section*{Data availability}
All the data that support the plots and the other findings of this study are available from the corresponding author upon reasonable request.

% \bibliography{bibliography}% Produces the bibliography via BibTeX.

\begin{thebibliography}{50}
\bibitem{note3}
\begin{botherref}
By thermalization, we mean the process of reaching thermal equilibrium, through
  equipartition of energy and uniform temperature that maximizes the system
  entropy. Since our system is not connected to a thermal bath, thermal
  equilibrium is obtained via nonlinear interactions.
\end{botherref}
\endbibitem
\end{thebibliography}

%%%
%% BioMed_Central_Bib_Style_v1.01

%%%

\newpage

\section*{Author contributions}
N.K.E. and D.N.C. conceived the project. N.K.E. performed the analytic calculations, the numerical simulations, and made all the plots. N.K.E. and D.N.C. wrote the paper. 

\section*{Competing interests}
The authors declare that there are no competing interests.
\newpage

\newpage

\newpage
\begin{figure}
\centerline{
\includegraphics[width=0.8\columnwidth]{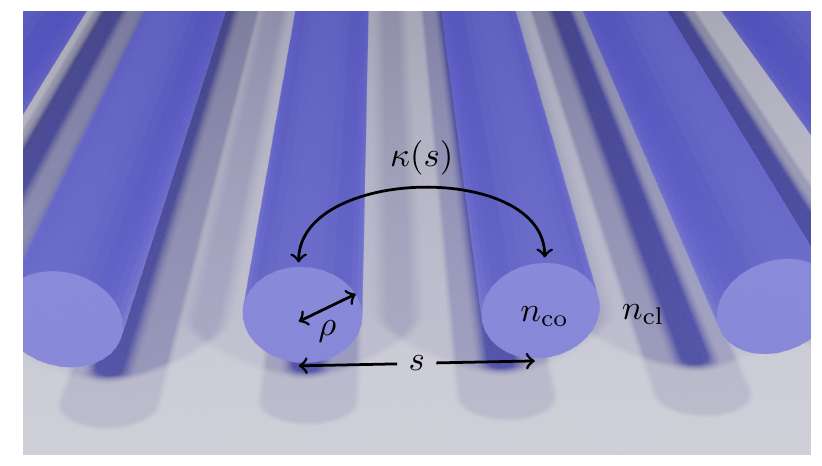}
}
\caption{Parameters of waveguide arrays. The radius of each waveguide is $\rho$, whereas the refractive index of the core and the cladding are $n_\mathrm{co}$ and $n_\mathrm{cl}$, respectively. The spacing between waveguides is $s$ and the coupling coefficient between adjacent waveguides is $\kappa(s)$.}
\label{fig:sup1}
\end{figure}
\section*{Supplementary Note: Coupling between weakly guided circular waveguides}
We consider an array of circular waveguides that are weakly guided. Each core has index $n_\mathrm{co}$ while $n_\mathrm{cl}$ is the refractive index of the surrounding fluid or cladding (see Fig.~\ref{fig:sup1}). The radius of each core is $\rho$, and the distance between successive elements of the lattice is $s$. The index contrast between the core and the cladding is
\[
  \Delta = \frac12\left(
    1-\frac{n_\mathrm{cl}^2}{n_\mathrm{co}^2}
      \right)
    \approx
    \frac{n_\mathrm{co}-n_\mathrm{cl}}{n_\mathrm{co}},
\]
%where $n_\mathrm{co}$, $n_\mathrm{cl}$ is the refractive index of the core and the cladding, respectively.
The fiber $V$ number is
\begin{equation}
  V = k_0\rho(n_\mathrm{co}^2-n_\mathrm{cl}^2)^{1/2},
\end{equation}
with $V^2=U^2+W^2$, 
\begin{equation}
  U = \rho (k_0^2n_\mathrm{co}^2-\beta^2)^{1/2},\quad
  W = \rho(\beta^2-k_0^2n_\mathrm{cl}^2)^{1/2},
\end{equation}
where
% $\rho$ is the radius of the core,
$k_0=2\pi/\lambda_0$, $\lambda_0$ is the free-space wavelength, and $\beta$ is the propagation constant. 
For a weakly guiding fiber, the eigenvalue problem for the HE$_{11}$  (LP$_{01}$) mode is obtained by requiring the mode and its derivative to be continuous
\begin{equation}
  U \frac{J_1(U)}{J_0(U)}
  =
  W \frac{K_1(U)}{K_0(U)}. 
\end{equation}
The coupling coefficient between two circular waveguides is given by~\cite{snyde-owt-1983}
\begin{equation}
  \kappa =
  \frac{\sqrt{2\Delta}}{\rho}
  \frac{U^2}{V^3}
  \frac{K_0(Ws/\rho)}{K_1^2(W)}, 
\end{equation}
an expression that can be very well approximated by
\begin{equation}
  \kappa \approx
  \left(
    \frac{\pi\Delta}{Wd\rho}
  \right)^{1/2}
  \frac{U^2}{V^3}
  \frac{e^{-Ws/\rho}}{K_1^2(W)}. 
\end{equation}

\renewcommand{\bibsection}{\vskip12pt\noindent\textbf{Supplementary References}}

\end{document}